# Metadata in the BioSample Online Repository are Impaired by Numerous Anomalies

Rafael S. Gonçalves, Martin J. O'Connor, Marcos Martínez-Romero, John Graybeal and Mark A. Musen

Stanford Center for Biomedical Informatics Research
Stanford University, CA, USA
`rafael.goncalves@stanford.edu`

**Abstract.** The metadata about scientific experiments are crucial for finding, reproducing, and reusing the data that the metadata describe. We present a study of the quality of the metadata stored in BioSample—a repository of metadata about samples used in biomedical experiments managed by the U.S. National Center for Biomedical Technology Information (NCBI). We tested whether 6.6 million BioSample metadata records are populated with values that fulfill the stated requirements for such values. Our study revealed multiple anomalies in the analyzed metadata. The BioSample metadata field names and their values are not standardized or controlled—15% of the metadata fields use field names not specified in the BioSample data dictionary. Only 9 out of 452 BioSample-specified fields ordinarily require ontology terms as values, and the quality of these controlled fields is better than that of uncontrolled ones, as even simple binary or numeric fields are often populated with inadequate values of different data types (e.g., only 27% of Boolean values are valid). Overall, the metadata in BioSample reveal that there is a lack of principled mechanisms to enforce and validate metadata requirements. The aberrancies in the metadata are likely to impede search and secondary use of the associated datasets.

## 1 Motivation

Metadata are key components of online data repositories and are essential for finding, retrieving, and reusing scientific data. Finding relevant scientific data requires not only that the data simply be accompanied by metadata, but also that the metadata be of sufficient quality for the corresponding datasets to be discovered and reused. When the quality of the metadata is poor, software systems that index and avail themselves of the experiment data may not find and return search results that otherwise would be appropriate for given search criteria. To have a chance at making use of poor-quality metadata would require a significant metadata post-processing effort. A frequently estimated figure for the cleanup time is approximately 80% of a scientist's data-analysis time [1]. The literature on metadata quality generally point to the need for better practices and infrastructure for authoring metadata. Bruce *et al.* [2] define various metadata quality metrics, such as completeness (e.g., all necessary fields should be filled in), accuracy (e.g., the values filled in should be specified as appropriate for the field), and provenance (e.g., information about the metadata author) of metadata. Park *et al.* [3, 4] surveyed multiple works on metadata quality, and specified several high-level principles for the creation of good-quality metadata. The metrics mentioned in these works have been recently supplemented by the FAIR data principles [5]. The FAIR principles specify desirable criteria that metadata and their corresponding datasets should meet to be Findable, Accessible, Interoperable, and Reusable.

Several empirical studies suggest that metadata need drastic improvement. Scarce use of ontologies to control field names and values and lack of validation of metadata fields have been identified as significant problems. For example, Zaveri [6] discovered that the metadata records in the Gene Expression Omnibus (GEO) suffer from redundancy, inconsistency, and incompleteness. This problem occurs because GEO allows users to create arbitrary fields that are not predefined by the GEO data dictionary. These fields and their values are not validated by GEO. Park [7] examined the use of Dublin Core (DC) elements in the field names of metadata records in a corpus of 659 records. Park identified various problems with the representation of DC elements, which could have been prevented with better infrastructure to map metadata field names to DC elements. Bui *et al.* [8] conducted a similar study to investigate the use of DC elements in metadata fields in a larger corpus of around 1 million records. The authors found that 6 DC fields are "rather well populated," while the other 10 fields that they analyzed were poorly populated on average. None of these authors, however, investigated whether the content of metadata fields is appropriately specified according to the fields' expected values. For example, the studies we mentioned checked whether DC fields are populated, but not whether the values for the *dc:date* field are actual dates formatted according to some identifiable standard, or whether the values for the language field resolve to controlled terms from an ontology about languages, or a language value set. For data to be FAIR, the value of each metadata field needs to be accurate and uniform (e.g., relying on controlled terms where possible), and to adhere to the field specification. Using controlled terms as a means to standardize metadata field names and field values allows users to be able to find data in a principled way, without having to cater to *ad hoc* representation mechanisms.

In this paper, we present an analysis of the quality of metadata in the BioSample metadata repository [9], maintained by the U.S. National Center for Biotechnology Information (NCBI). BioSample stores metadata that describe the biological materials (samples) under investigation in a wide range of projects. Our goal is to measure the quality of the metadata records in BioSample based on the fields that these records specify. We consider metadata to be of good quality if the metadata fields are controlled, and if the expected field values are complied with. We analyzed BioSample fields (so-called *attributes*) that have computationally-verifiable expectations for their values. An attribute is a pair consisting of an attribute name and an attribute value. For example, values for the attribute named *disease* of human samples should correspond to terms in the Human Disease Ontology (DOID). We used the BioPortal [10] search service to find correspondences between metadata values and ontology terms. In all, we analyzed 73 of the 452 attributes that BioSample specifies.

## 2 Materials and Methods

We acquired a copy of the BioSample database from the central NCBI FTP archive on June 25$^{th}$, 2017.[1] The BioSample database is distributed as an XML file, with no explicit versioning information. Our copy of BioSample contained 6,615,347 metadata records.

We built software infrastructure to extract key bits of information about each metadata record in the BioSample database, and to test whether the attributes of each sample record were filled in and well-specified.[2] The tool can be used to repeat the procedure of verifying the validity of BioSample records. Our tool collects the following data: sample identifier, accession, publication date, last update date, submission date, access (public or controlled), identifier and name of the sample organism, owner name, package name, status (live or suppressed), and status date. Then, for each attribute within our tested attributes, the software records the attribute name, its value, and verifies whether it is filled in according to the attribute's specification. An attribute specification describes the format and content of the expected attribute value.

**BioSample overview**
Officially launched in 2011, BioSample accepts submissions of metadata through a Web-based portal, which guides users through a series of metadata-entry forms. The first form is for choosing a *package*. A package represents a type of sample and it specifies a set of attributes that should be used to describe samples of that particular type. For instance, the Human package requires its records to have the attributes: *age*; *sex*; *tissue*; *biomaterial provider*; and *isolate*. The Human package lists other attributes that can be optionally provided. Each of the 104 BioSample package types has a different set of rules regarding which attributes are required or are optional.[3] The Generic package is a package type that has no requirements at all. This package is not listed in the package documentation, and it is not an option in BioSample's Web forms. The Generic records in BioSample were either harvested from other repositories or submitted through BioSample's less-stringent programmatic interfaces.

BioSample provides a dictionary of 452 metadata attribute names that can be used to describe the samples that form the substrates of experiments.[4] Metadata authors can, however, provide additional attributes with arbitrary names with no guidance or control from BioSample. Each metadata record describing a sample can contain multiple attributes. Given the domain, we expect BioSample metadata to use terms from ontologies in BioPortal—a repository that currently hosts 585 biomedical ontologies.

**Analysis of BioSample metadata**
Our study assesses the quality of metadata in BioSample according to whether the attributes in the metadata records specify (1) a controlled attribute name (i.e., provided by an ontology), (2) an attribute name that is in BioSample's attribute dictionary, and (3) a valid value according to the attribute specification (Fig. 1). We analyzed all the

---

[1] Obtained from: http://ftp.ncbi.nih.gov/biosample/biosample_set.xml.gz.
[2] The software is open-source: http://github.com/metadatacenter/biosample-analyzer.
[3] Full list of packages available at: https://www.ncbi.nlm.nih.gov/biosample/docs/packages.
[4] Full list of attributes available at: https://www.ncbi.nlm.nih.gov/biosample/docs/attributes.

BioSample attributes and categorized those attributes that have the same type of expected values into the following groups:

**Ontology-term attributes**. There are 9 BioSample attributes that take on term values from specific ontologies. For example, the attribute *phenotype*, representing the phenotype of the sampled organism, should have input values that are terms from the Phenotypic Quality Ontology (PATO). To verify whether ontology terms supply values for attributes in BioSample when appropriate, we performed searches in BioPortal for exact matches of the possible values for each relevant BioSample attribute field within the ontology that the BioSample attribute documentation indicates should provide values for that field. We indicate that an ontology-term attribute is *well-specified* if its value matches a term in the designated ontology.

**Term attributes.** There are 8 attributes in BioSample packages that require terms as values. Unlike the ontology-term group of attributes, these 8 attributes are not linked to a specific ontology that should be used to provide the values. Given this ambiguity, we attempt to verify the values assigned to these attributes by performing a search in BioPortal for *any* matching term, but we do not require an exact ontology match, and we do not restrict the search to any specific set of ontologies. A term attribute is *well-specified* if its value matches a term in some ontology in BioPortal.

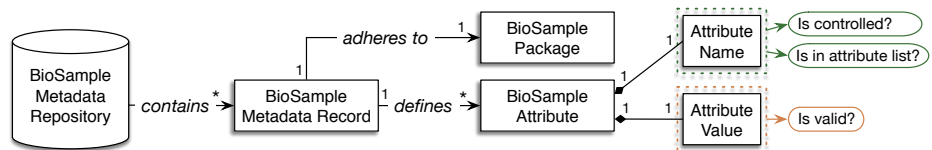

**Fig. 1.** Overview of BioSample. The BioSample repository contains metadata records that adhere to packages. Each metadata record defines multiple attributes. An attribute is composed of an attribute name and a value. We aim to determine whether attribute names are standardized and in BioSample's attribute list, and whether attribute values are valid.

**Value set attributes.** There are 32 attributes whose values are constrained to value sets specified in the BioSample documentation. We programmed methods for verifying that values stored for each of these types of attributes are appropriate, and we tested whether the values found in BioSample records actually corresponded to the values defined in the BioSample documentation.

**Boolean attributes.** We tested 4 attributes in BioSample packages that require a Boolean value. We indicate that a Boolean attribute is *well-specified* if its value is *true* or *false*, regardless of capitalization. We consider values such as *f* or *yes* to be invalid.

**Integer attributes.** We tested 4 attributes that require an integer value. An integer attribute is *well-specified* if the given value can be parsed into an integer.

We gathered similar information about other structured attributes, although we did not test the validity of those values in the BioSample data. For example, there are 161 attributes that require a unit, 21 attributes that require a PubMed ID, and so on.

We chose to validate the 5 groups above because the characteristics of these groups are easily tested, and because the expected values of the attributes are straightforward for users to specify (e.g., compared to attributes such as those that require a value to be composed of a floating-point number followed by a special symbol).

## 3 Results

The sample records in BioSample represent 94 unique package types. Thus, not all of the 104 BioSample packages types are used. Generic packages make up for the bulk of the BioSample database—85% of the records use this package definition. The next most populated package is *Pathogen*, consisting of 3.2% of the records.

The metadata in BioSample typically are rich with attributes (i.e., name-value pairs). BioSample records contain a total of 82,360,966 attributes—either attributes using names specified by the BioSample dictionary or attributes with custom names. A total of 12,284,229 attributes (15% of all attributes), spread over 63,354 metadata records (0.96% of all records), use attribute names that are not specified in the BioSample attribute dictionary. In these attributes, we identified 18,198 unique custom attribute names specified by submitters. The records that contain these attributes have been submitted by 313 different laboratories. Overall there are 18,650 attribute names used in BioSample metadata records—452 are BioSample-specified (2.4%), and the remainder are user-specified (97.6%). Only 9 of the 452 BioSample attributes have values that are standardized using ontologies. None of BioSample's attribute names seem to be controlled. Of all BioSample records, only 197,123 Generic-package records (0.03%) do not specify any attributes. On average, each BioSample metadata record specifies 12 attributes. The vast majority of BioSample records (97%) specify at least one attribute.

The main results of our study are presented in Table 1. We will now explain each of the rows in the table as they appear.

**Table 1.** BioSample attributes according to their type. The columns show the number of attributes that are filled-in with values, and those whose values are well-specified.

| Attribute type | # Filled-in values | # Well-specified values | % Well-specified values |
|---|---|---|---|
| Ontology term | 1,976,642 | 639,154 | 32% |
| Value set | 4,165,320 | 3,842,733 | 92% |
| Boolean | 7,585 | 2,015 | 27% |
| Integer | 163,535 | 120,701 | 74% |

**Ontology-term attributes**. Most ontology-term attributes in BioSample contain poorly-specified values rather than terms from ontologies. There are 1,016,483 records (15.4%) that contain a value for one or more attributes that ideally require an ontology term. Out of those, only 43% (441,719) of the records have valid values for their ontology term attributes. These records specify a total of 1,976,642 ontology term attributes, and only 639,154 of them (32%) are actually filled in with ontology terms. Some values for these attributes do not match with terms in BioPortal because they are not typed correctly, or contain strange symbols. For example, the *disease* attribute requires a term from the Human Disease Ontology (DOID), and some values given include *gastrointestinal stromal tumor_4* (*gastrointestinal stromal tumor* is a class in DOID), *HIV_Positive* (*HIV* is a class in DOID), *infected with Tomato spotted wilt virus isolate p105RBMar*, which does not have a close match, *lung_squamous_carcinoma* that would have matched with a term if not for the underscores, numeric values that do not match BioPortal terms, and so on.

**Term attributes**. Our analysis of the term attributes group revealed a vast, highly-customized value space for these uncontrolled attribute values. There are attribute values that clearly have no correspondence with controlled terms in BioPortal ontologies. For example: *for pig and horse*, which yields a first match with *Horse spray and rub-on*; *hive cell S29*, which yields a first match with *Hive Island*; numbers; entire sentences; etc. To properly determine the quality of these values they would require additional post-processing, or a different ontology term matching strategy. We plan to analyze this group of attributes at greater depth in future work.

**Value set attributes**. The attributes that use value sets are the most well-specified in BioSample from among of the attribute groups we analyzed. There are 4,028,758 records that contain one or more value set attributes. Out of those, 3,781,283 records (94%) contain value set attributes that appear to be valid. These records specify a total of 4,165,320 value set attributes, and 3,842,733 (92%) of those are well-specified. Even though most records adhere to the value sets, we observed that a wide range of values is given for even such simple attributes of a sample as *sex*. This attribute has possible values: *male*, *female*, *pooled male and female*, *neuter*, *hermaphrodite*, *intersex*, *not determined*, *missing*, *not applicable*, and *not collected*. The values in BioSample records include variations of the accepted values *male* and *female* such as *m*, *f*, *Male*, *FEMALE*, etc. Other values include: *pool of 10 animals, random age and gender*; *juvenile*; *Sexual equality*; *parthenogenic*; *larvae, pupae and adult (queens - workers)*; *castrated horse*; *gynoparae*; *uncertainty*; *Vaccine and Infectious Disease Division*; *Clones arrayed from a variety of cDNA libraries*; *Department I of Internal Medicine*; some numeric values; values containing only symbols; misspelled words such as *mal e*, *makle*, *femLE*, etc.

**Boolean attributes**. The Boolean attributes are the worst quality of all the attributes we analyzed. Overall, 6,767 BioSample records contain a value for one or more Boolean-type attributes. Only 2,013 (30%) of those records have attributes that are valid. These records specify 7,585 Boolean-type attributes, of which only 2,015 (27%) are well-specified. For example, for the *smoker* attribute, there are such diverse values as: *Non-smoker*, *nonsmoker*, *non smoker*, *ex-smoker*, *Ex smoker*, *smoker*, *Yes*, *No*, *former-smoker*, *Former*, *current smoker*, *Y*, *N*, *0*, *--*, *never*, *never smoker*, among others.

**Integer attributes**. The values for integer attributes are mostly well-specified. There are 158,854 records containing one or more attributes that require an integer value. Out of those, 120,026 records (76%) contain valid attributes. These records specify a total of 163,535 integer attributes, and 120,701 of those (74%) are well-specified. The attribute *medication code*[5] does not have a single valid value (e.g., one of the individual values is *Insulin glargine injectable solution; Insulin lispro injectable solution; Fluoxetine; Simvastatin; Isosorbide mononitrate; Amlodipine/Omelsartan medoxomil;*). The attribute *host taxonomy ID*, which should be filled in with integer values from the NCBI taxonomy,[6] has such input values as: *e;N/A, Mus musculus, NO;* etc.

---

[5] BioSample documentation does not specify why this attribute should take numeric values.

[6] The expected integer values for this attribute are identifiers that should correspond to organism names in the NCBI organismal classification.

## 4     Conclusions

We carried out a survey of the quality of metadata in the BioSample repository by verifying the validity of several collections of attributes. Our study revealed multiple, significant anomalies in the metadata hosted in the BioSample repository. While BioSample makes available specialized packages to provide some control over metadata submissions, the clear majority of submitters prefer to use the (restriction-free) Generic package, which has no controls or requirements. A significant proportion of the attributes (15%) in BioSample records use ad hoc attribute names that do not exist in BioSample's attribute dictionary. These 18,198 custom attribute names account for the large majority of attribute names (97.6%) used in metadata records, signaling a need to go beyond the 452 attribute names specified by BioSample. A considerable number of ontology-term attributes (68%) have values that do not correspond to actual ontology terms. The Boolean-type attributes have a staggeringly wide range of values, and only 27% of them are valid according to the BioSample documentation.

Our results demonstrate that the use of ontologies in BioSample metadata is rather immature, even though BioSample is a relatively modern initiative that aims at encouraging the standardization of metadata using controlled terms. Although the requirements for BioSample metadata are well-specified, these do not seem to be enforced during metadata authoring. The result is clear: We observed that the metadata in BioSample are generally of poor quality, and that the underlying repository suffers from a lack of appropriate infrastructure to enforce metadata requirements. The use of ontology terms is particularly substandard, and even simple fields that require Boolean or integer values are often populated with nonparsable values.

While the quality of the BioSample attribute groups that are standardized (i.e., those that require terms from ontologies or value sets) is disappointing, it is not as poor as the non-controlled attributes such as the Boolean attributes. This situation suggests that metadata should make far more use of controlled terms, particularly since there is such a wide range of well-established biomedical ontologies available in BioPortal.

In general, our study suggests that there is a need for a more robust approach to authoring metadata. To be FAIR, metadata should be represented using a formal knowledge representation language, and they should use ontologies that follow FAIR principles to control the metadata attributes. These aspects help to ensure interoperability of the metadata, and are crucial for finding data based on their metadata. The metadata should follow a principled representation of metadata and metadata fields based on Web standards. The tooling available to scientists who author metadata should impose appropriate restrictions on the metadata. For example, where a value should be a term from a specific ontology, the metadata author should only be presented with options that are valid terms from that ontology when filling in metadata.

These findings guide our implementation of a software system that aims to transform the way scientists author metadata to ensure standardization, completeness, and consistency. The Center for Expanded Data Annotation and Retrieval (CEDAR) [11] is developing a suite of tools—the CEDAR Workbench [12–14]—which allows users to build metadata templates based on community standards, to fill in those templates with metadata values that are appropriately authenticated, to upload data and their metadata to online repositories, and to search for metadata and templates stored in the CEDAR repository. The goal of CEDAR is to improve significantly the quality of the metadata submitted to public repositories, and thus to make online scientific datasets more FAIR.

In future work, we will expand our survey of the quality of metadata in BioSample to cover all the attributes that are computationally verifiable using automated methods. We also intend to carry out similar analyses of other online metadata repositories.

**Acknowledgements**

This work is supported by grant U54 AI117925 awarded by the National Institute of Allergy and Infectious Diseases through funds provided by the trans-NIH Big Data to Knowledge (BD2K) initiative (http://www.bd2k.nih.gov). NCBO is supported by the NIH Common Fund under grant U54HG004028.